\documentclass[12pt, prd, showpacs]{revtex4}
\usepackage{amssymb}
\usepackage{amsmath}

\input{tcilatex}

\begin{document}

\title{Is the super-Penrose process possible near black holes?}
\author{O. B. Zaslavskii}
\affiliation{Department of Physics and Technology, Kharkov V.N. Karazin National
University, 4 Svoboda Square, Kharkov 61022, Ukraine}
\affiliation{Institute of Mathematics and Mechanics, Kazan Federal University, 18
Kremlyovskaya St., Kazan 420008, Russia}
\email{zaslav@ukr.net }

\begin{abstract}
We consider collisions of particles near generic axially symmetric extremal
black holes. We examine possibility of indefinitely large extraction of
energy (the so-called super-Penrose process) in the limit when the point of
collision approaches the horizon. Three potential options are considered
(fractional powers of the lapse function in the relations between the
energies and the angular momenta of particles in the point of collision),
collision between outgoing particles and ingoing ones, collision in the
ergoregion far from the horizon). It turns out in all three cases that
states suitable for the super-Penrose process cannot be obtained from the
previous collision of particles with finite masses and angular momenta.
\end{abstract}

\maketitle

\section{Introduction}

Collisions of two particles near a black hole can lead to unbound energy $%
E_{c.m.}$ in their centre of mass, provided one of them is fine-tuned. This
observation made in Ref. \cite{ban} for extremal Kerr black holes,
stimulated significantly interest to the energetics of particle collisions
in strong gravitational field. At first, the main part of researches
concentrated on the question for which objects and under which condition
large $E_{c.m.}$ are possible thus extending the results of \cite{ban} to
more general types of spacetimes and scenarios. Meanwhile, there is another
question which is under active discussion just now. Whether or not it is
possible to gain unbound Killing energy $E$ of debris measured at infinity
after collision and thus obtain unbound energy extraction? In several works
it was shown that for collisions near black holes, in spite of unbound $%
E_{c.m.}$, the energy $E$ turns out to be quite modest. This was shown for
the Kerr metric in \cite{p}, \cite{j} and arbitrary "dirty" (surrounded by
matter) black holes in \cite{z}. These results were obtained for collisions
between ingoing particles sent to a black hole from infinity.

Meanwhile, there exist more involved scenarios. It was noticed in \cite%
{shnit} numerically that collision between fine-tuned outgoing particle and
ingoing not fine-tuned ("usual") one in the Kerr background leads to rather
significant amplification of the energy $E$. This result was confirmed
analytically in \cite{pir152}, \cite{pn}. Moreover, it was noticed in \cite%
{card} numerically and in \cite{mod} analytically that head-on collision
between two usual particles near black holes gives rise to unbound $E$ (it
was called super-Penrose process in \cite{card}). The problem, however,
consists in that it is difficult to create an outgoing usual particle in the
immediate vicinity of a black hole where all particles with finite energy
should be ingoing. One can try to get a usual particle near the horizon as a
result of previous collision but it turns out that for finite masses and
angular momenta of initial particles this is impossible not only for the
Kerr metric \cite{pir15} but in much more general case \cite{epl1}.

However, previous studies of collisions near black holes do not exhaust all
possible cases since there are three additional potential options for
getting unbound $E$ not explored yet. (i) One can try to weaken requirement
on the type of particles and consider the case intermediate between usual
and fine-tuned ("critical") particles. Namely, one can probe such a
deviation from the special relation between the energy and angular momentum
that includes fractional powers of the lapse function at the point of
collision. (ii) The conclusion about serious (with the factor about 14) but
still restricted amplification of the energy due to collision was obtained
for the Kerr metric only. It is of interest to consider generic axially
symmetric stationary black holes, not specifying their metric and to
elucidate, whether or not the unbound $E$ are possible. (iii) There are
scenarios in which collision occurs not near the horizon but somewhere in
the ergoregion \cite{ergo}, \cite{mergo}, provided the angular momentum of
either particle is large negative. It was shown quite recently that the
super-Penrose process for such scenario is possible by itself \cite{sergo}.
However, the question arises, how to prepare a state with initial large
angular momentum. More precisely, can it be obtained in the preceding
collision with all finite characteristics of particles?

In the present work, we discuss all three aforementioned issues. It turns
out that in all three cases the answer is negative that gives new
restrictions on physical realization of the super-Penrose process (although
it is not excluded in principle). \ We assume that there is no interaction
between particles and the electromagnetic field. Throughout the paper, we
put fundamental constants $G=c=1$.

\section{Basic equations}

We consider the metric%
\begin{equation}
ds^{2}=-N^{2}dt^{2}+g_{\phi }(d\phi -\omega dt)^{2}+\frac{dr^{2}}{A}%
+g_{\theta }d\theta ^{2}\text{,}  \label{met}
\end{equation}%
in which the coefficients do not depend on $t$ and $\phi $. Correspondingly,
the energy $E$ and angular momentum $L$ are conserved. In the present work,
we will be interested in the equatorial motion. Then, the geodesic equation
of motion gives us 
\begin{equation}
m\frac{dt}{d\tau }=\frac{X}{N^{2}},
\end{equation}%
\begin{equation}
m\frac{d\phi }{d\tau }=\frac{L}{g_{g}}+\frac{\omega X}{N^{2}},  \label{phi}
\end{equation}%
where $m$ is the mass, $X=E-\omega L$ and and $\tau $ is the proper time.
The forward-in-time condition reads $\frac{dt}{d\tau }>0$, whence%
\begin{equation}
X>0  \label{x}
\end{equation}%
everywhere for $N\neq 0$.

We assume that there exists the horizon on which $N=0$. We assume that the
horizon is extremal. If $X_{H}=0$ (that is compatible with the
forward-in-time condition), we call a particle critical (hereafter,
subscript "H" means that the corresponding quantity is taken on the
horizon). If $X_{H}\neq 0$ and is separated from zero, we call a particle
usual. In what follows, we will need further classification. We call a
particle near-critical if $X_{H}=O(N_{c})$ and, for brevity, fractional if 
\begin{equation}
X_{H}=O(N_{c}^{s})  \label{fr}
\end{equation}%
with 
\begin{equation}
0<s<1\text{.}  \label{s}
\end{equation}%
Hereafter, subscript "c" denotes quantities calculated at the point of
collision.

We restrict ourselves by motion within the plane $\theta =\frac{\pi }{2}$
where we scale the radial coordinate to achieve $A=N^{2}$. Using equation (%
\ref{phi}) and normalization condition for the four-velocity, one can obtain
that the radial momentum%
\begin{equation}
p^{r}=\sigma Z\text{,}  \label{prz}
\end{equation}%
where

\begin{equation}
Z=\sqrt{X^{2}-\alpha ^{2}}\text{,}  \label{z}
\end{equation}%
\begin{equation}
\alpha ^{2}=N^{2}\delta ^{2}\text{, }\delta ^{2}=m^{2}+\frac{L^{2}}{g_{\phi }%
}\text{,}  \label{alpha}
\end{equation}%
$\sigma =-1$ for the ingoing particle and $+1$ for the outgoing one.

Let two particles 1 and 2 collide to produce particles 3 and 4. Then, the
conservation laws at the point of collision give us for the energy and
angular momentum:%
\begin{equation}
E_{1}+E_{2}=E_{3}+E_{4}\text{,}  \label{E}
\end{equation}%
\begin{equation}
L_{1}+L_{2}=L_{3}+L_{4}\text{.}  \label{L}
\end{equation}%
It follows from (\ref{E}), (\ref{L}) that%
\begin{equation}
X_{1}+X_{2}=X_{3}+X_{4}\text{.}  \label{x12}
\end{equation}%
The conservation of the radial momentum reads%
\begin{equation}
p_{1}^{r}+p_{2}^{r}=p_{3}^{r}+p_{4}^{r}\text{.}  \label{p}
\end{equation}

In view of (\ref{prz}), (\ref{z}), it is equivalent to 
\begin{equation}
\sigma _{1}Z_{1}+\sigma _{2}Z_{2}=\sigma _{3}Z_{3}+\sigma _{4}Z_{4}\text{.}
\label{pz}
\end{equation}

We assume that it is particle 3 that escapes to infinity.

\section{Fractional dependence on $N$}

We deal with a black (not white) hole, so a typical particle that moves
along the complete geodesics with finite nonzero energy cannot move away
from a black hole in the immediate vicinity of the horizon (see below for
details). However, if a particle is originated from some previous collision,
this cannot be excluded in advance. Let such a particle ($\sigma _{1}=+1$)
collide with the second one having $\sigma _{2}=-1$ (head-on collision). Let
us consider the simplest situation when both particles have the same energy $%
E_{0},$ mass $m$ and angular momentum $L_{0}$. It is also supposed that
products of collisions have the same mass $\mu $. Then, it was found in \cite%
{rot2} that the maximum possible energy of debris

\begin{equation}
E_{\max }=E_{0}+\frac{(\omega X_{0}+\frac{L_{0}N^{2}}{g_{\phi }})\sqrt{%
g_{\phi }}}{N}\sqrt{B}\text{, }B=1-\frac{\mu ^{2}N^{2}}{X_{0}^{2}-\frac{%
N^{2}L_{0}^{2}}{g_{\phi }}}\text{,}  \label{e+}
\end{equation}%
where all quantities are taken at the point of collision. Eq. (\ref{e+}) was
obtained in the context not connected with black holes. However, it is exact
and is valid independently of the presence or absence of the horizon.

Let collision occur very nearly to the horizon, so $N\rightarrow 0.$ If both
particles are usual, $X_{H}\neq 0$, so $E_{\max }$ grows like $N^{-1}$ and
the energy extraction from a black hole is unbound. This is in agreement
with the previous numeric \cite{card} and analytical \cite{mod}
observations. However, the problem is that there exist severe restrictions
on such a scenario and, for finite masses and angular momenta, it is
impossible to create a usual particles in previous collisions \cite{pir15}, 
\cite{epl1}.

To find way out, one can try to weaken requirement on $X$ and allow
fractional particles. In general, near the extremal horizon, the following
expansion is valid (see \cite{dirty} for details) near the horizon:%
\begin{equation}
X=X_{H}+B_{1}LN+O(N^{2})\text{,}  \label{xn}
\end{equation}%
where $B_{1}>0$ is a constant. For fractional particles, it is the first
term in (\ref{xn}) which dominates at the collision point due to (\ref{s}).
Then, it follows from (\ref{e+}) that $E_{\max }\sim N_{c}^{s-1}$. It still
diverges, although more slowly than for usual particles.

Meanwhile, as is said above, such a scenario includes the outgoing particle
in the immediate vicinity of the horizon that, in turn, requires preceding
collision in which this particle was created. The most physically
interesting state corresponds to two particles falling form infinity. At
present, it is already known that in this case, the collision between a
usual and near-critical particles near the horizon gives rise to a
near-critical particle \cite{p} - \cite{z}, so for both particles 1 and 3
the quantity $X=O(N_{c})$. However, the case of fractional particles was not
considered before. Below, we fill this gap and elucidate, whether or not it
is possible to create the outgoing fractional particle in collision of two
ones moving from infinity.

We consider such a scenario in which (i) particles 1 and 2 come from
infinity, so $\sigma _{1}=\sigma _{2}=-1$, (ii) particle 3 escapes to
infinity. We imply that masses and angular momenta of initial particles are
finite, so $X_{1}$ and $X_{2}$ are finite as well. Since $X>0$, $X_{3}$ and $%
X_{4}$ are also finite. We can choose at our will the finite quantities $%
\left( X_{1}\right) _{c}$ and $\left( X_{2}\right) _{c}$. It is obvious from
(\ref{pz}) that it is impossible to have $\sigma _{3}=\sigma _{4}=+1,$ so at
least one of particles 3 or 4 should have negative $\sigma $. We assume that 
$\sigma _{4}=-1$ but retain $\sigma _{3}=\pm 1$. This allows to take into
account the scenario in which particle 3 moves immediately after collision
towards the horizon and bounces back.

Thus we have%
\begin{equation}
-\sqrt{X_{1}^{2}-\alpha _{1}^{2}}-\sqrt{X_{2}^{2}-\alpha _{2}^{2}}=\sigma
_{3}\sqrt{X_{3}^{2}-\alpha _{3}^{2}}-\sqrt{X_{4}^{2}-\alpha _{4}^{2}}\text{.}
\label{1234}
\end{equation}

Now, different scenarios with fractional particles are considered
separately. In doing so, we somewhat modify the approach developed in \cite%
{j} for the Kerr metric and in \cite{z} for dirty black holes.

\subsection{Particle 1 is fractional, particle 2 is usual}

Now, 
\begin{equation}
\left( X_{1}\right) _{c}=\beta _{1}N_{c}^{q}\text{,}  \label{1f}
\end{equation}%
where $0<q<1$, the constant $\beta _{1}>0$.

Then, at the point of collision $\alpha _{1}^{2}=O(N_{c}^{2})$, 
\begin{equation}
Z_{1}\approx X_{1}-\frac{\alpha _{1}^{2}}{2X_{1}}\approx X_{1}-CN^{2-q}\text{%
.}  \label{z1}
\end{equation}%
\ Here $C=\frac{1}{2\beta _{1}}(\frac{L_{1}^{2}}{g_{\phi }}+m_{1}^{2})$. It
follows from (\ref{x}) and (\ref{1234}) that particle 4 is usual and%
\begin{equation}
X_{3}+\sigma _{3}\sqrt{X_{3}^{2}-\alpha _{3}^{2}}\approx CN_{c}^{2-q}.
\end{equation}

If $\sigma _{3}=+1$,%
\begin{equation}
X_{3}=O(N_{c}^{2-q})=O(N_{c}^{s})\text{,}  \label{x3s}
\end{equation}%
\begin{equation}
\alpha _{3}=O(N_{c}^{2-q})=O(N_{c}^{s})\text{, }  \label{al}
\end{equation}%
where 
\begin{equation}
s=2-q>1\text{ }
\end{equation}%
in contradiction with (\ref{s}). Thus this scenario is unsuitable for our
goal.

If $\sigma _{3}=-1$, there are two options. If $\alpha _{3}\lesssim X_{3}$,
estimates (\ref{x3s}), (\ref{al}) are still valid and the scenario should be
rejected as well. If $\alpha _{3}\ll X_{3}$, there is no turning point for
particle 3, i falls into a black hole and cannot escape. Again, the scenario
is unsuitable for our purposes.

\subsection{Particle 1 is fractional, particle 2 is near-critical}

In the case under discussion,%
\begin{equation}
\left( X_{1}\right) _{c}=\beta _{1}N_{c}^{s_{1}}\text{, }s_{1}<1\text{,}
\end{equation}%
\begin{equation}
\left( X_{2}\right) _{c}=\beta _{2}N_{c}\text{,}  \label{be2}
\end{equation}%
where $\beta _{1}$ and $\beta _{2}$ are positive constants. Now, $\left(
X_{2}\right) _{c}\ll \left( X_{1}\right) _{c}$ and gives negligible
contribution into $X$. Then, the conservation law entails that at least one
of particles has $X$ of the order $N_{c}^{s_{1}}$. Let%
\begin{equation}
\left( X_{3}\right) _{c}\approx \beta _{3}N_{c}^{s_{1}}\text{.}  \label{b3}
\end{equation}%
Then, eq. (\ref{x12}) entails%
\begin{equation}
\left( X_{4}\right) _{c}\approx (\beta _{1}-\beta _{3})N_{c}^{s_{1}}\text{, }%
\beta _{1}>\beta _{3}\text{.}  \label{b4}
\end{equation}%
As $Z_{3.4}^{2}>0$, we have%
\begin{equation}
\alpha _{3,4}=\tilde{\alpha}_{3,4}N_{c}^{s_{1}}\text{,}  \label{a34}
\end{equation}%
where $\tilde{\alpha}_{3,4}$ are finite and, in general, nonzero.

Taking into account that $s_{1}<1$, $2-s_{1}>1$, we can omit the second term
in the expansion (\ref{z1}). We also neglect $Z_{2}$ in the conservation law
(\ref{1234}) since it has the order $N_{c}$, whereas $Z_{1}$ has the order $%
N_{c}^{s_{1}}$. Therefore, we have%
\begin{equation}
-\beta _{1}N_{c}^{s_{1}}\approx \sigma _{3}\sqrt{X_{3}^{2}-\alpha _{3}^{2}}-%
\sqrt{X_{4}^{2}-\alpha _{4}^{2}}\text{.}  \label{bea}
\end{equation}%
By substitution of (\ref{b3}) - (\ref{a34}), we obtain 
\begin{equation}
\beta _{1}=-\sigma _{3}\sqrt{\beta _{3}^{2}-\tilde{\alpha}_{3}^{2}}+\sqrt{%
(\beta _{1}-\beta _{3})^{2}-\tilde{\alpha}_{4}^{2}}\equiv D.  \label{ba}
\end{equation}

Let $\sigma _{3}=+1.$ If $\tilde{\alpha}_{3}=\tilde{\alpha}_{4}=0$, $D=\beta
_{1}-2\beta _{3}<\beta _{1}$, so eq. (\ref{ba}) is contradictory. If $\tilde{%
\alpha}_{3}\neq 0$ or $\tilde{\alpha}_{4}\neq 0$, 
\begin{equation}
D<\sqrt{\beta _{3}^{2}-\tilde{\alpha}_{3}^{2}}+\sqrt{(\beta _{1}-\beta
_{3})^{2}-\tilde{\alpha}_{4}^{2}}<\beta _{3}+\beta _{1}-\beta _{3}=\beta _{1}%
\text{,}
\end{equation}%
again in contradiction with (\ref{ba}). \ 

For $\sigma _{3}=-1$, eq. (\ref{ba}) can \ hold, provided $\tilde{\alpha}%
_{3}=\tilde{\alpha}_{4}=0$. \ But if $\tilde{\alpha}_{3}=0$, there is no
turning point for particle 3 and it falls down to a black hole and cannot
escape.

Thus the present scenario is unsuitable for our purposes.

\subsection{Particle 1 is fractional, particle 2 is fractional}

Now we have%
\begin{equation}
\left( X_{1}\right) _{c}=\beta _{1}N_{c}^{s_{1}}\text{,}
\end{equation}%
\begin{equation}
\left( X_{2}\right) _{c}=\beta _{2}N_{c}^{s_{2}}.
\end{equation}%
In the expansion (\ref{z1}) (with $q$ replaced with $s$), the first and
second terms coming from $Z_{1}$ have the order $N_{c}^{s_{1}}$ and $%
N_{c}^{2-s_{1}}$, respectively. The similar terms in $Z_{2}$ have the order $%
N_{c}^{s_{2}}$ and $N_{c}^{2-s_{2}}$. Let, for definiteness, $s_{1}<s_{2}$.
Then, the conservation law (\ref{1234}) entails in the main approximation
just eq. (\ref{bea}). It follows from the conservation law (\ref{x12}) that 
\begin{equation}
X_{3}+X_{4}=O(N_{c}^{s_{1}})\text{. }
\end{equation}%
As $X_{1,2}>0$ due to the forward-in-time condition (\ref{x}), each of them
has, in general, the same order. Then,%
\begin{equation}
\alpha _{3.4}\leq X_{3,4}\approx \beta _{3,4}N_{c}^{s_{1}}\text{,}
\end{equation}%
$\beta _{3}$ and $\beta _{4}$ are positive constants. Then, the main
formulas from the previous scenario apply here with the same conclusion.

If $s_{1}=s_{2}$, one should take into account in (\ref{1234}) contributions
from both particles 1 and 2. Therefore, instead of $\beta _{1}$, the
combination $\beta _{1}+\beta _{2}$ appears that can be denoted as a new
constant $\beta $. Further, the previous formulas again apply. Thus this
does not affect the conclusion.

\section{Kinematic explanation}

Here, we discuss qualitatively, what are the underlying kinematic reason
that create obstacles to trajectories with usual \cite{epl1} and fractional
outgoing particles.

\subsection{Individual particle}

For any particle moving in the axially symmetric stationary background,
there exists the relation between the energy and angular momentum \cite{k}
at the given point with $N=N_{c}$ :%
\begin{equation}
X=N_{c}E_{loc}\text{,}
\end{equation}%
where $E_{loc}$ is the energy measured by the local observer with the zero
angular momentum (ZAMO \cite{72}),%
\begin{equation}
E_{loc}=\frac{m}{\sqrt{1-V^{2}}}\text{,}
\end{equation}%
$V$ is the local velocity in the ZAMO frame.

Another physical characteristics, important in this context, is the proper
time for motion between the point in the vicinity of the horizon where $%
N=N_{c}$ and the point with a given $N$ outside the horizon. It follows from
(\ref{prz}) that%
\begin{equation}
\tau =m\int_{r_{c}}^{r}\frac{dr}{Z}\text{,}  \label{time}
\end{equation}%
where $N(r_{c})=N_{c}$. The horizon lies at $r=r_{+}$ where $N(r_{+})=0$. In
the present work we consider extremal horizons for which $N\sim r-r_{+}$.
For nonextremal ones, $N^{2}\sim r-r_{+}$.

For any finite $E_{loc},$ one obtains that $\lim_{N_{c}\rightarrow
0}X=X_{H}=0$, so the particle becomes critical. By contrary, if $X_{H}\neq 0$
(a usual particle), it entails%
\begin{equation}
\lim_{N_{c}\rightarrow 0}E_{loc}=\infty \text{,}
\end{equation}%
so this energy becomes infinite, a particle approaches the horizon with the
speed of light.

If such a particle moves in the inward direction, it reaches the horizon in
a finite proper time both for nonextremal and extremal horizons since for
usual particles $Z\neq 0$ on the horizon. This is quite typical situation
that occurs, in particular, for the Schwarzschild metric.

Let now a usual particle be outgoing. Then, it means that it crossed the
horizon from inside a finite proper time ago and keeps moving further away
from the horizon. But this is the situation of a white, not a black hole and
should be rejected in the present context.

This difficulty does not arise if a particle has $V<1$ separated from zero.
As the proper distance to the extremal horizon is infinite, the proper time
between the horizon and any other point outside a black hole is now infinite
(it diverges as $\,-\ln N_{c}$ with $N_{c}\rightarrow 0$). This happens if $%
X_{H}=0$, so the particle is critical. Such a particle could not arrive from
the inner region and there is no contradiction with the nature of a black
hole.

If a particle is fractional, it is seen from (\ref{z}), (\ref{fr}), (\ref%
{time}) that the proper time for traveling between a given point with $%
r=r_{c}$ and the horizon $r=r_{+}$ is finite since although $X_{H}$ is
small, it foes not vanish. The same reasoning as in the case of usual
particles applies here. Therefore, such individual particles cannot move in
the vicinity of black (not white) holes in the outward direction.

\subsection{Restriction on particles created from collision and behavior of $%
E_{c.m.}$}

If a particle appears as a result of collision, its trajectory starts just
at the point of collision and cannot be extended into past. Therefore, the
arguments based on the continuity of geodesics are not valid anymore.
Kinematic reasonings cannot, in general, replace the detailed analysis of
bookkeeping of energy and momenta. Meanwhile, it is also instructive to look
at the problem in another way, considering the behavior of the energy in the
centre of mass frame $E_{c.m.}.$

Let two particles 1 and 2 fall from infinity, collide and produce particles
3 and 4. Then, 
\begin{equation}
E_{c.m.}^{2}=-(m_{1}u_{1\mu }+m_{2}u_{2\mu })((m_{1}u_{1}^{\mu
}+m_{2}u_{2}^{\mu })=-(m_{3}u_{3\mu }+m_{3}u_{3\mu })((m_{4}u_{4}^{\mu
}+m_{4}u_{4}^{\mu })\text{,}
\end{equation}%
\begin{equation}
E_{c.m.}^{2}=m_{1}^{2}+m_{2}^{2}+2m_{1}m_{2}\gamma _{12}\text{,}  \label{12}
\end{equation}%
where $\gamma =-u_{1\mu }u^{2\mu }$. One can deduce from equations of motion
that \cite{prd}%
\begin{equation}
m_{1}m_{2}\gamma _{12}=\frac{X_{1}X_{2}-Z_{1}Z_{2}}{N^{2}}-\frac{L_{1}L_{2}}{%
g_{\phi }}  \label{ga12}
\end{equation}%
In a similar way, the same energy can be expressed in terms of
characteristics of particles 3 and 4. Assuming that after collisions they
move in opposite directions, one obtains that%
\begin{equation}
E_{c.m.}^{2}=m_{3}^{2}+m_{4}^{2}+2m_{3}m_{4}\gamma _{34}\text{,}  \label{34}
\end{equation}%
\begin{equation}
m_{3}m_{4}\gamma _{34}=\frac{X_{3}X_{4}+Z_{3}Z_{4}}{N^{2}}-\frac{L_{3}L_{4}}{%
g_{\phi }}\text{.}
\end{equation}%
Using the conservation laws (\ref{x12}), (\ref{p}) one can check that both
expressions for $E_{c.m.}$ coincide. Below we restrict ourselves by
considering two cases. (The others can be considered in a similar way.)

\subsection{All particles are usual}

If all particles are usual, eq. (\ref{12}) gives us that for given masses, $%
\lim_{N\rightarrow 0}E_{c.m.}^{2}$ is finite, whereas eq. (\ref{34}) entails
that it diverges like $N^{-2}$. This contradiction shows that the process
under discussion is impossible for finite parameters of reaction.

However, the reaction becomes possible if one adjusts the mass of initial
particle (say, $m_{1}$) adjusted to $N_{c}$ and allows it to diverge. Let $%
E_{1}=m_{1}$, $m_{3}$ be finite. The conservation of energy gives us that $%
X_{4}\approx E_{4}\approx m_{1}$. Then, for large $m_{1}$ one obtains from (%
\ref{12})%
\begin{equation}
E_{c.m.}^{2}\approx m_{1}^{2}\text{.}
\end{equation}

It follows from (\ref{34}) that%
\begin{equation}
E_{c.m.}^{2}\approx 4\frac{m_{1}X_{3}}{N_{c}^{2}}\text{.}
\end{equation}%
Comparing both expressions, one finds%
\begin{equation}
m_{1}\approx \frac{4X_{3}}{N_{c}^{2}}
\end{equation}%
in agreement with \cite{pir15} and \cite{epl1}.

For the same reasons, two usual particles with finite masses falling from
infinity cannot produce a near-critical outgoing particle.

\subsection{Collision of usual and fractional particles}

Let us try collision between a usual and fractional particles falling from
infinity. Equations (\ref{12}), (\ref{ga12}) and (\ref{z1}) entail that%
\begin{equation}
E_{c.m.}^{2}=O(N_{c}^{-s})\text{.}  \label{1s}
\end{equation}%
Now, we expect that particle 4 is usual and particle 3 is fractional, $%
X_{3}=O(N_{c}^{q})$, $\alpha =O(N_{c}^{q}),$ the value of $q$ is found
below. Eq. (\ref{34}) gives rise to%
\begin{equation}
E_{c.m.}^{2}=O(N_{c}^{q-2}).  \label{3q}
\end{equation}

Comparing (\ref{1s}) and (\ref{3q}) we see that $q=2-s$. But in view of (\ref%
{s}), we obtain that $q>1$, so the attempt to find a scenario is
unsuccessful. Thus evaluations of the energy in the centre of mass frame
agrees with the analysis of the conservation law for the radial momentum.

\section{Head-on collision with participation of near-critical particle}

It follows from the results above that it is impossible to create a
fractional particle in preceding collision between two particles falling
from infinity. Thus we return to the known situation when it is the
near-critical particle that escapes to infinity after the first collision.
But one may ask, whether or not its further collision with particles of
other types can lead to the unbound energy. Now, let outgoing near-critical
particle 1 collide with ingoing particle 2 and produces particle 3 that
escapes to infinity. We assume that $\sigma _{4}=-1$, so particle 4 falls
into a black hole.

The conservation law (\ref{p}) reads

\begin{equation}
\sqrt{X_{1}^{2}-\alpha _{1}^{2}}-\sqrt{X_{2}^{2}-\alpha _{2}^{2}}=\sqrt{%
X_{3}^{2}-\alpha _{3}^{2}}-\sqrt{X_{4}^{2}-\alpha _{4}^{2}}  \label{pr}
\end{equation}

We want to elucidate, whether or not we can obtain unbound energy $E_{3}$.
If $E_{3}\rightarrow \infty $ but $X_{3}$ is finite, it follows from (\ref{x}%
) that%
\begin{equation}
L_{3}\rightarrow \infty \text{.}  \label{inf}
\end{equation}

To answer our question, we consider three cases separately.

\subsection{Particle 2 is usual}

For small $N_{c}$,%
\begin{equation}
\left( X_{1}\right) _{c}\approx A_{1}N_{c}\text{, }  \label{a1}
\end{equation}%
\begin{equation}
\left( X_{2}\right) _{c}=O(1)\text{,}  \label{a2}
\end{equation}%
$A_{1}>0$ is a constant. It follows from (\ref{x12}) that one of produced
particle is usual and another one is near-critical. Neglecting terms of the
order $N_{c}^{2}$, we obtain from (\ref{pr}) that it is particle 3 which is
near-critical, while particle 4 is usual and at the point of collision,%
\begin{equation}
A_{1}N-X_{2}\approx -X_{4}+\sqrt{X_{3}^{2}-\alpha _{3}^{2}}\text{.}
\end{equation}%
Taking into account (\ref{xn}) we obtain%
\begin{equation}
(X_{4})_{c}\approx (X_{2})_{c}+O(N_{c})\text{,}
\end{equation}%
and%
\begin{equation}
\sqrt{\left( X_{3}^{2}\right) _{c}-\alpha _{3}^{2}}=O(N_{c})\text{. }
\end{equation}%
In turn, this entails that 
\begin{equation}
\alpha _{3}\leq \left( X_{3}\right) _{c}=O(N_{c})\text{.}  \label{ax}
\end{equation}%
Then, it follows from (\ref{alpha}) that $L_{3}$ is finite. This is in
contradiction with (\ref{inf}), so we cannot obtain unbound $E_{3}$.

\subsection{Particle 2 is near-critical}

Then, in addition to condition (\ref{a1}), we have the similar one $\left(
X_{2}\right) _{c}\approx A_{2}N_{c}$ instead of (\ref{a2}). The conservation
law (\ref{pr}) gives us that also $\left( X_{3}\right) _{c}\approx
A_{3}N_{c} $ and $\left( X_{4}\right) _{c}\approx A_{4}N_{c}$ $\ $with
constants $A_{3}$ and $A_{4}$. Then, using (\ref{ax}) we obtain that $L_{3}$
is finite in contradiction with (\ref{inf}), so this scenario is
incompatible with the super-Penrose process.

\subsection{Particle 2 is fractional}

Now, particle 1 is near-critical, particle 2 is fractional. This is
completely similar to the situation considered above. Now,%
\begin{equation}
\left( X_{1}\right) _{c}=\beta _{1}N_{c}\text{, }\left( X_{2}\right)
_{c}=\beta _{2}N_{c}^{q}\text{, }0<q<1,  \label{x2}
\end{equation}%
\begin{equation}
\left( X_{3}\right) _{c}\approx \beta _{3}N_{c}^{q}\text{, }\left(
X_{4}\right) _{c}\approx (\beta _{2}-\beta _{3})N_{c}^{q}.
\end{equation}

Then, eq. (\ref{pr}) gives us%
\begin{equation}
-\beta _{2}N_{c}^{q}=\sigma _{3}\sqrt{\left( X_{3}\right) _{c}^{2}-\alpha
_{3}^{2}}-\sqrt{\text{ }\left( X_{4}\right) _{c}^{2}-\alpha _{4}^{2}}.
\end{equation}%
This equation coincides with (\ref{bea}) if $\beta _{1}$ is replaced with $%
\beta _{2}$. The same analysis as before applies with the conclusion that
the scenario is forbidden.

\subsection{When particles 3 and 4 move in the same direction}

In this Subsection, we discuss the case $\sigma _{4}=\sigma _{3}$. This is
compatible with the general type of scenario under discussion. Let $\sigma
_{3}=+1$, so both particles 3 and 4 escape.

Then,we have from (\ref{pz}) that%
\begin{equation}
\sqrt{X_{1}^{2}-\alpha _{1}^{2}}-\sqrt{X_{2}^{2}-\alpha _{2}^{2}}=\sqrt{%
X_{3}^{2}-\alpha _{3}^{2}}+\sqrt{X_{4}^{2}-\alpha _{4}^{2}}.  \label{dir}
\end{equation}

Here, particle 1 is near-critical by assumption. If particle 2 is usual we
obtain obvious contradiction in the limit $N_{c}\rightarrow 0$. If particle
2 is fractional, its contribution still dominates for small $N_{c}$, and the
contradiction arises again. Thus particle 2 should be near-critical. Then,
it follows from (\ref{dir}) that particles 3 and 4 are also near-critical.

Let us write $X_{i}=x_{i}N$, $\alpha _{i}=\delta _{i}N$ ($1\leq i\leq 4$).
Then, we have 
\begin{equation}
\sqrt{x_{1}^{2}-\delta _{1}^{2}}-\sqrt{x_{2}^{2}-\delta _{2}^{2}}=\sqrt{%
x_{3}^{2}-\delta _{3}^{2}}+\sqrt{x_{4}^{2}-\delta _{4}^{2}},  \label{x1}
\end{equation}%
where all coefficients have the order $O(1)$.

According to (\ref{alpha}), this entails that $L_{.3}$ and $L_{4}$ are
finite. This is inconsistent with (\ref{inf}), so that $E_{3}$ is finite and
the super-Penrose process is impossible.

The case $\sigma _{3}=\sigma _{4}=-1$ can be considered in a similar way
with the same conclusion.

\section{Collisions inside ergoregion}

In Ref. \cite{ergo} another scenario was suggested for the Kerr metric
(generalized in \cite{mergo} for a generic metric (\ref{met})). Two
particles collide in the ergoregion but not in the immediate vicinity of the
horizon. One of them has the large negative angular momentum. Then, the
energy in the centre of mass grows indefinitely. It was shown recently \cite%
{sergo} that in this process $E$ can be made as large as one likes, so the
super-Penrose process is possible. However, the question remains how to
create a particle with indefinitely large $\left\vert L\right\vert .$ The
most attractive option is to get it from some previous collision as was
suggested in \cite{ergo}. Below we examine this possibility and show that
this is impossible.

Let two particles 1 and 2 with finite $E_{1,2}$ and $L_{1,2}$ collide. We
assume that particle 4 has $L_{4}<0$ with $\left\vert L_{4}\right\vert
\rightarrow \infty $. Due to the conservation of energy and momentum, $%
L_{3}\rightarrow +\infty $. Meanwhile, $X_{3}$ and $X_{4}$ remain finite
since they are both positive and obey (\ref{x12}) where $X_{1}$ and $X_{2}$
are positive. In turn, this means that the condition $Z^{2}>0$ with $Z$
taken from (\ref{z}) cannot be satisfied for particles 3 and 4 since $X$ is
finite, while $L^{2}\rightarrow \infty $.

\section{Remarks about nonequatorial motion}

Throughout the paper we considered equatorial motion only. More general case
of nonequatorial motion is beyond the scope of the present paper, so we
restrict ourselves by brief remarks. We expect no qualitative changes in
this case for the following reasons. (i) In that case, in (\ref{alpha}) one
should simply replace $m^{2}$ with $m^{2}+m^{2}g_{\theta }\left( \frac{%
d\theta }{d\tau }\right) ^{2}$. In particular, near the horizon, where $%
N\rightarrow 0$, this changes the coefficients in the expansion (\ref{z1})
but does not affect general dependence of $Z$ on $N_{c}$, so previous
analysis of the conservation of the radial momentum applies. (iii) In
general, the analysis gets more complicated since one should take into
account the $\theta -$component of the momenta. However, this additional
conditions can only restrict further the possibility of the super-Penrose
process, so our negative results retain their validity.

\section{Conclusion}

In the present work, we filled some gaps left from previous consideration.
Its results in combination with those of the previous one \cite{epl1} give a
kind of a no-go theorem that applies to the collisional Penrose process.
Namely, either (i) the state of initial particles does not lead to the
super-Penrose process at all or (ii) it formally leads to it but such a
state cannot be realized near a black hole from initial conditions with
finite masses and angular momenta.

However, it does not forbid large (but restricted) extraction whose value is
model-dependent. Say, for the Kerr metric the corresponding factor is about
14 \cite{shnit} - \cite{pn}.

One can also think of creating states, suitable for the super-Penrose
process, not due to mechanical collisions but, say, from statistical
fluctuations in thermal gas surrounding a black hole.

\begin{acknowledgments}
This work was funded by the subsidy allocated to Kazan Federal University
for the state assignment in the sphere of scientific activities.
\end{acknowledgments}


\begin{thebibliography}{99}
\bibitem{ban} M. Ba\~{n}ados, J. Silk and S.M. West, Phys. Rev. Lett. 
\textbf{103} (2009) 111102 [arXiv:0909.0169].

\bibitem{p} M. Bejger, T. Piran, M. Abramowicz, and F. H\aa kanson, Phys.
Rev. Lett.\textbf{\ 109} (2012) 121101 [arXiv:1205.4350].

\bibitem{j} T. Harada, H. Nemoto and U. Miyamoto, Phys. Rev\textit{. }D%
\textbf{\ 86} (2012) 024027 [Erratum ibid. D 86 (2012) 069902]
[arXiv:1205.7088].

\bibitem{z} O. Zaslavskii, Phys. Rev. D\textbf{\ 86} (2012) 084030
[arXiv:1205.4410].

\bibitem{shnit} J. D. Schnittman, Phys. Rev. Lett. \textbf{113}, 261102
(2014), [arXiv:1410.6446].

\bibitem{pir152} E. Leiderschneider and T. Piran, [arXiv:1510.06764].

\bibitem{pn} K. Ogasawara, T. Harada, and U. Miyamoto [arXiv:1511.0011].

\bibitem{card} E. Berti, R. Brito and V. Cardoso, Phys.\ Rev.\ Lett. \textbf{%
114}, 251103 (2015) [arXiv:1410.8534].

\bibitem{mod} O. Zaslavskii, Mod. Phys. Lett. A 30 (2015) 1550076,
[arXiv:1411.0267].

\bibitem{pir15} E. Leiderschneider and T. Piran, [arXiv:1501.01984].

\bibitem{epl1} O. B. Zaslavskii, Europhys. Lett. 111 (2015) 50004,
[arXiv:1506.06527].

\bibitem{ergo} A. A. Grib and Yu. V. Pavlov, Europhys. Lett. 101, 20004
(2013) [arXiv:1301.0698].

\bibitem{mergo} O. B. Zaslavskii, Mod. Phys. Lett. A. Vol. 28, 1350037
(2013) [arXiv:1301.4699].

\bibitem{sergo} O.B. Zaslavskii, [arXiv:1510.02140].

\bibitem{rot2} O. B. Zaslavskii, [arXiv:1511.00844].

\bibitem{dirty} I. V. Tanatarov and O. B. Zaslavskii, Phys. Rev. D \textbf{86%
}, 044019 (2012) [arXiv:1206.2580].

\bibitem{k} O. B. Zaslavskii, Phys. Rev\textit{.} D \textbf{84}, 024007
(2011) [arXiv:1104.4802].

\bibitem{72} J. M. Bardeen, W. H. Press, and S. A. Teukolsky, Astrophys. J. 
\textbf{178}, 347 (1972).

\bibitem{prd} O.B. Zaslavskii, Phys. Rev. D \textbf{82} (2010) 083004
[arXiv:1007.3678].
\end{thebibliography}
\end{document}